\documentclass{article}

\PassOptionsToPackage{numbers, compress}{natbib}

\usepackage[final]{neurips_2020}

\usepackage[utf8]{inputenc} 
\usepackage[T1]{fontenc}    
\usepackage{url}            
\usepackage{booktabs}       
\usepackage{amsfonts}       
\usepackage{nicefrac}       
\usepackage{microtype}      
\usepackage{xcolor}
\usepackage[numbib]{tocbibind}

\usepackage{graphicx}
\graphicspath{ {./figures/} }
\title{Keyword spotting - Detecting commands in speech using deep learning}

\author{
  Sumedha Rai \\
  \texttt{sr5387@nyu.edu} \\
  Center for Data Science \\
  New York University \\
  \And
    Tong Li \\
   \texttt{tl2204@nyu.edu} \\
  Center for Data Science \\
  New York University \\
  \And
  Bella Lyu \\
  \texttt{hl4229@nyu.edu} \\
  Center for Data Science \\
  New York University \\
   }
  
\begin{document}

\maketitle

\begin{abstract}
  Speech recognition has become an important task in the development of machine learning and artificial intelligence. In this study, we explore the important task of keyword spotting using speech recognition machine learning and deep learning techniques. We implement feature engineering by converting raw waveforms to Mel Frequency Cepstral Coefficients (MFCCs), which we use as inputs to our models. We experiment with several different algorithms such as Hidden Markov Model with Gaussian Mixture, Convolutional Neural Networks and variants of Recurrent Neural Networks including Long Short-Term Memory and the Attention mechanism. In our experiments, RNN with BiLSTM and Attention achieves the best performance with an accuracy of 93.9 \%.


\end{abstract}

\section{Introduction}

Thanks to the rapid development of speech recognition systems, interacting with technology using voice has become commonplace. Human-computer interaction has now become a critical multidisciplinary field of study and speech commands are used in a wide variety of customer-centric environments to enable easy execution of tasks. \\
\\
Keyword spotting is a branch of speech recognition that deals with the identification of special speech commands or ‘keywords’ in utterances. With a small footprint size, keyword spotting has a wide scope of applications, such as robotics, virtual assistants, and vehicle mounted electronics. \\
\\
In the field of retail, using ‘keywords’ can enable a customer to get updates on orders or be directed to the customer help centers with simple keyword responses such as ‘yes’ and ‘no’. Virtual assistants use a variety of small special speech commands such as ‘Hey Siri’ or ‘Alexa’ to switch from sleep mode to wake up mode and respond to requests by the user. In the field of robotics, service robots can be directed to move in specific directions such as ‘left’, ‘right’, ‘backward’, ‘forward’ or execute actions using other keywords. 
For keyword spotting to be efficient, it is important that we achieve a good accuracy of speech recognition, low-latency and be able to run in computationally constrained environments (such as mobile devices).\\
\\
Most fields of speech recognition are usually approached with knowledge applied in time series analysis because of its sequential input feature of the sound wave. For this study, we aim to develop several models on an existing public dataset, the Google Speech Commands Dataset, to achieve the task of recognizing keywords in speech by taking its sequential nature into account.\\ \\
We use the more traditional Hidden Markov Model with Gaussian Mixture (HMMGMM) as our baseline, and expand our algorithms to deep neural nets. We find that deep neural nets outperform the HMMGMM significantly. Among deep neural nets, RNN with BiLSTM and Attention achieves the highest accuracy. 

\section{Related Work}

\label{headings}



Hidden Markov Models have been widely used for speech recognition as a traditional time series model (Juang et al. (1991) \cite{hmm-prev}). While recently, deep learning algorithms are considered as more powerful and have been widely implemented and applied in speech recognition. Among the deep neural architectures, convolutional neural networks have been frequently used (Hamid et al. (2014) \cite{cnn-prev}). Graves et al. (2013) \cite{6638947} also adopt Long Short-term Memory RNNs for speech recognition. A combined architecture named convolutional recurrent network with attention was later introduced by de Andrade et al. (2018) \cite{deandrade2018neural} for keyword spotting on 5 different tasks. More complex architectures have also been developed to recognize speech in English and Mandarin that can also efficiently deal with noises and accents (Amodei et al. (2015) \cite{amodei2015deep}). 


\section{Problem Definition and Algorithms}
\subsection{Task}

In our experiments, for HMMGMM and RNNs, the inputs will be the Mel-Frequency Cepstral Coefficients (MFCCs) with 12 components. For CNN, since it is already an effective feature extractor, instead of using the MFCCs, we resample the file with a new rate of 8000Hz instead of 16000Hz as in the original audio as we notice a large portion of each utterance is in silence, and we use this resampled audio as our inputs for CNN. The outputs will be the uttered commands from the speech. To be more specific, we have a total of 35 commands of single words like "on" and "off", and we convert these words to integers ranging from 0 to 34. Our task thus becomes building different algorithms to classify the speech into one of the 35 categories, and compare their results.  
    

    

\subsection{Algorithms}
\subsubsection{Hidden Markov Model with Gaussian Mixture (HMMGMM)}
We use the HMMGMM as our baseline model. In this model, we start by specifying the $N$ hidden states, and $M$ distinct observations per state. We also have the transition probability matrix $A$, where each $A_{ij}$ represents the probability of transitioning from $state_i$ to $state_j$, and an emission matrix $B$ where each $B_i(k)$ represents the probability of generating observation $k$ at state $j$. In the model, it is drawn from the Gaussian Mixture distribution. We also have an initial distribution $\pi$ where each $\pi_i$ denotes the probability of starting at $state_i$. In our experiment, we choose six hidden states and two GMM mixtures as our final model configuration as it gives the best validation set result.

Let $\theta = \{A, B, \pi\}$ represent the parameters and $\mathbf{z}$ denote the hidden states sequence. In the model, we wish to maximize $$P(x|\theta) = \int_\mathbf{z} P(x, \mathbf{z} | \theta) \textbf{dz}$$ by using Expectation-Maximization Algorithm (EM). To do this, we start with the E-step by maximizing $$q(\mathbf{z}) = P(\mathbf{z} | x, \theta^{old}) \: \textrm{with fixed parameter} \: \theta^{old}$$ 
We then implement the M-step by maximizing $$Q(\theta, \theta^{old}) = \int_z P(\mathbf{z} | x, \theta^{old}) log P(x, \mathbf{z} | \theta) \: \textrm{while holding} \: P(\mathbf{z} | x, \theta^{old}) \: \textrm{fixed}$$

\subsubsection{Convolutional Neural Networks (CNN)}
We also experiment with deep neural nets by starting with CNN. In this model, the sequential nature is not taken into account and the data is treated as a bag-of-words style. We use the M5 CNN structure \cite{m5}, as illustrated in Table 1. In this model, the convolution operator slides through the input matrix as a square kernel with different sizes into numerous channels, which acts as a feature extractor, and a batch normalization is applied. Then a max pooling is used to reduce the size and pass it to the next convolutional layer. Finally it goes through a fully connected layer and softmax for prediction.  

We use learning rate of 0.01 and Adam as our optimizer and the cross-entropy loss as our criteria. The model is trained in ten epochs.



\begin{table}
\begin{center}
\begin{tabular}{|c|}
\hline
input \\
\hline
Conv (kernel size 80) \\
BatchNorm \\
MaxPool (kernel size 4) \\
\hline
Conv (kernel size 3) \\
BatchNorm \\
MaxPool (kernel size 4) \\
\hline
Conv (kernel size 3) \\
BatchNorm \\
MaxPool (kernel size 4) \\
\hline
Conv (kernel size 3) \\
BatchNorm \\
MaxPool (kernel size 4) \\
\hline
Fully-Connected \\
\hline
Softmax \\
\hline

\end{tabular}
\end{center}
\label{tab:m5}
\caption{M5 CNN Structure}
\end{table}

\subsubsection{Recurrent Neural Networks}
\begin{figure} [!htpb]
    \centering
    \includegraphics[scale=0.8]{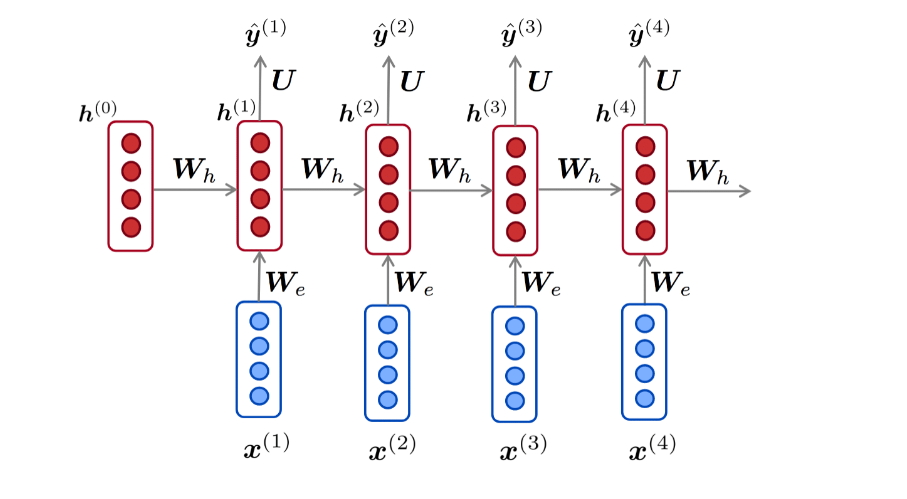}
    \caption{RNN Unrolled}
    \label{fig:rnn}
\end{figure}

We also adopt RNN for our task considering the sequential nature of the inputs. RNN is a form of deep neural nets that exhibit temporal dynamics. RNN differs from the CNN in that at each time step of the input sequence, a hidden state is computed by taking both the input and the previous hidden state into account, and this hidden state is then used to produce the output. Figure \ref{fig:rnn} illustrates this structure. Instead of the vanilla RNN cell, we adopt two variants of it. We start by building a RNN with unidirectional LSTM and modify this model by adding Bidirection LSTM cell and the attention mechanism.

\paragraph{Long Short-Term Memory}
The LSTM utilizes several gates to eschew the vanishing gradient problem and to allow a longer temporal dependency between inputs. The structure is well explained by \cite{snlp}. The first gate is the forget gate, which aims to delete information that is no longer needed from previous states, and then extract the information that we need from the previous state and current inputs, as illustrated below. $$f_t = \sigma(U_f h_{t-1} + W_f x_t)$$ $$k_t = c_{t-1} \odot f_t$$
$$g_t = tanh(U_g h_{t-1} + W_g x_t)$$
where $\sigma$ is the logistic sigmoid.

We then compute the add gate to add information to the current context. $$i_t = \sigma (U_i h_{t-1} + W_i x_t)$$ $$j_t = g_t \odot i_t$$ $$c_t = j_t + k_t $$ 

Finally we compute the output gate to decide what information is needed for the hidden state at the current time step. $$o_t = \sigma (U_o h_{t-1} + W_o x_t)$$ $$h_t = o_t \odot tanh(c_t)$$



\begin{figure} [!htpb]
    \centering
    \includegraphics[scale=0.15]{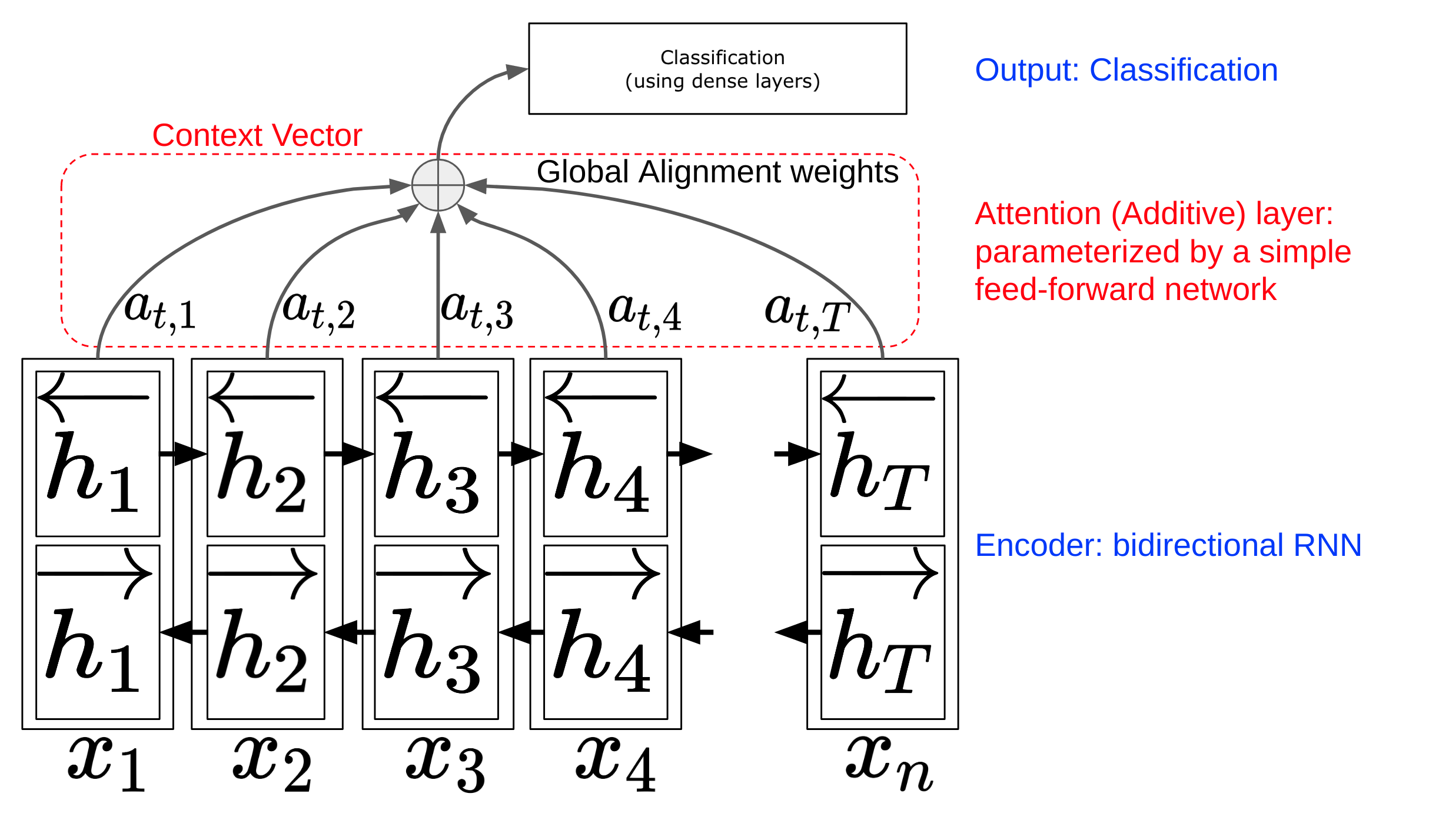}
    \caption{BiLSTM RNN with attention mechanism }
    \label{fig:attn}
\end{figure}

\paragraph{Attention RNN}
Recent work has shown that incorporating the attention mechanism in RNN architecture can significantly improve accuracy and is now an integral part of modern RNN frameworks.\\
\\
Our audio signal files contain keywords /speech commands that can potentially be anywhere within the 1 second length of the file, although it is expected to be centred within the audio file. Since our models are able to extract keywords and classify them, they should also be able to delineate the region where the speech exists. Thus, incorporating the attention mechanism in our RNN structure seems apposite for our task.\\
\\
In a standard model, only the last hidden state of the encoder RNN is used for output. Using an attention mechanism, we can retain and utilize all the hidden states from the input sequence. This will allow the model to extract information from all positions and focus on the relevant part of the input sequence by selectively picking out the region that has the speech command. Figure \ref{fig:attn} shows the architecture and input-output framework of a BiLSTM RNN with attention \cite{ateer2018}.

For our study, in addition to the models described above, we also adapt the attention RNN model as proposed by de Andrade et al. (2018) \cite{deandrade2018neural} for understanding the attention mechanism within a standard RNN model and for a comparative analysis of the results. The model uses convolutions and BiLSTM units followed by the attention layer composed of a dense layer for projection that is fed into a softmax function. The weighted average of the LSTM layers is fed into dense layers for classification. 

\section{Experimental Evaluation}
\subsection{Data}

We use the open-sourced Google Speech Commands Dataset v2 \cite{speechcommandsv2} for training and evaluating our models. Released in April 2018, this dataset consists of 105,829 one-second utterances of 35 words. 
The dataset has commands that can be used to build basic voice interfaces for use in different fields such as Internet of Things or robotics. Some examples are: 
    \begin{itemize}
        \item Common Commands: 'yes', 'no', 'on, 'off'
        \item Directions: 'left', 'right', 'up', 'down'
        \item Numbers: 'zero' to 'nine'
        \item Auxiliary words: 'bed', 'cat', 'marvin', 'wow'
    \end{itemize}
We assess our models on 35 word recognition task that tries to correctly predict each spoken word.

\subsubsection{Data Pre-processing}

\begin{figure} [!htpb]
    \centering
    \includegraphics[scale = 0.8]{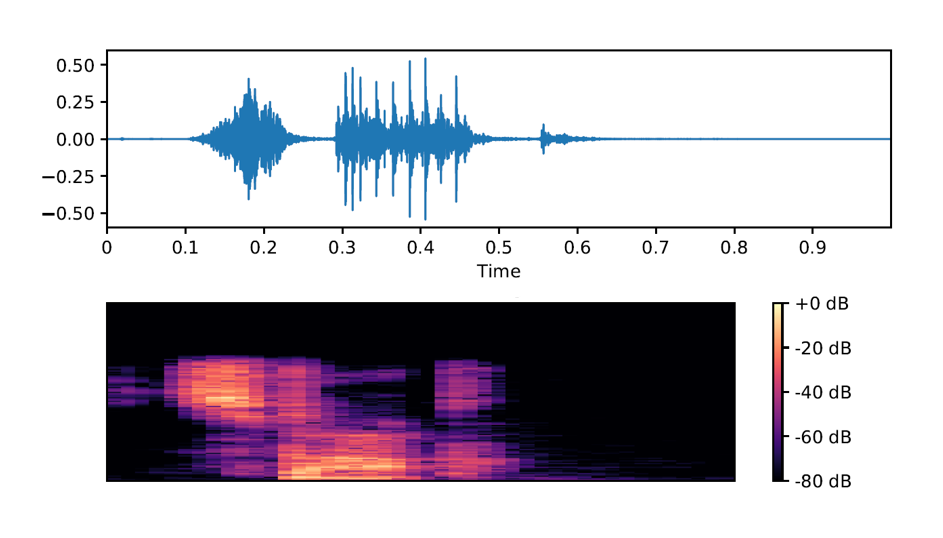}
    \caption{An Example of Waveform and MFCCs of "stop"}
    \label{fig:mf}
\end{figure}

To eliminate unnecessary information from the audio files, we start by extracting features that will be fed into the models. For this, we need to identify the components of the audio signal that represent the linguistic content of the audio and discard the part that carries unimportant information such as background noise. The non-linear Mel Scale relates perceived frequency, or pitch, of a pure tone to its actual measured frequency. Humans are much better at discerning small changes in pitch at low frequencies than they are at high frequencies. Thus, the Mel Frequency Cepstral Coefficients (MFCCs) are computed in a way to recognize the discrepancy of the human ear's critical bandwidths with frequency filters spaced linearly at low frequencies and logarithmically at high frequencies to retain the phonetically vital properties of a speech signal and accurately represents the phoneme being produced. They are the principal input features that have been widely used in automatic speech and speaker recognition and have been the gold standard in ASR since their introduction in the 1980s by Davis et al. \cite{1163420}. \\
We convert the audio files into 12 components MFCCs through the librosa package \footnote{https://librosa.org/} and then use this as model inputs for HMMGMM and RNNs. An example of the raw wave plot and converted MFCCs of the word "stop" is shown on Figure \ref{fig:mf}. For CNN, a resampled frequency of 8000Hz audio signal is used as the input. 

\subsection{Methodology}

Since the goal of our task is to enable the model to correctly identify the commands, we use accuracy as our metric for evaluation. The test accuracy is calculated for all model frameworks and reported together as a comparative analysis. Our experiment aims to investigate if deep neural nets have better performance than the more traditional time series analysis models, and if so, which architecture has the best performance. We split our data into training, validation and test sets with an 80-10-10 split.  


\subsection{Results}
\begin{figure}[!htpb]
    \centering
    \includegraphics[scale = 0.8]{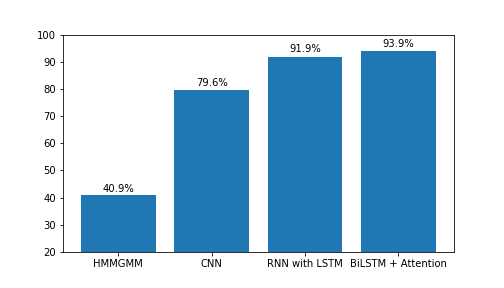}
    \caption{Model Accuracy}
    \label{fig:accs}
\end{figure}

The test set accuracy are shown in Figure \ref{fig:accs}. We notice that the deep neural nets outperform the traditional statistical model (HMMGMM) significantly. Among the deep nets, we also find that the two RNNs have higher test accuracy than the CNN, and the RNN with BiLSTM and attention achieves the best performance.

\subsection{Discussion}
By comparing the results, we find that deep architectures achieve higher accuracy. For CNN, it achieves an increase of 38.7 \% in the absolute scale of accuracy compared to HMMGMM. One potential reason for this is that the convolutional layers are effective in extracting features that are more important for our task. While for the three deep neural nets, both RNNs have substantial improvements in test set accuracy than the CNN (an improvement of 20.3 \% and 22.3 \% in absolute scale). This agrees with the assumption that RNNs are more powerful when the input is sequential in nature. In addition, we also notice an increase in performance when we add the BiLSTM and Attention mechanisms. However, this increase is not very significant (only 2\% increase in absolute scale). Some potential explanations include: 1. The LSTM already incorporates the previous states well with the forget and add gates, and thus an attention mechanism may not contribute much to the prediction. 2. The attention mechanism is possibly more appropriate for the encoder-decoder structure instead of the single-output RNN. 



\begin{figure} [!htpb]
    \centering
    \includegraphics[scale = 0.4]{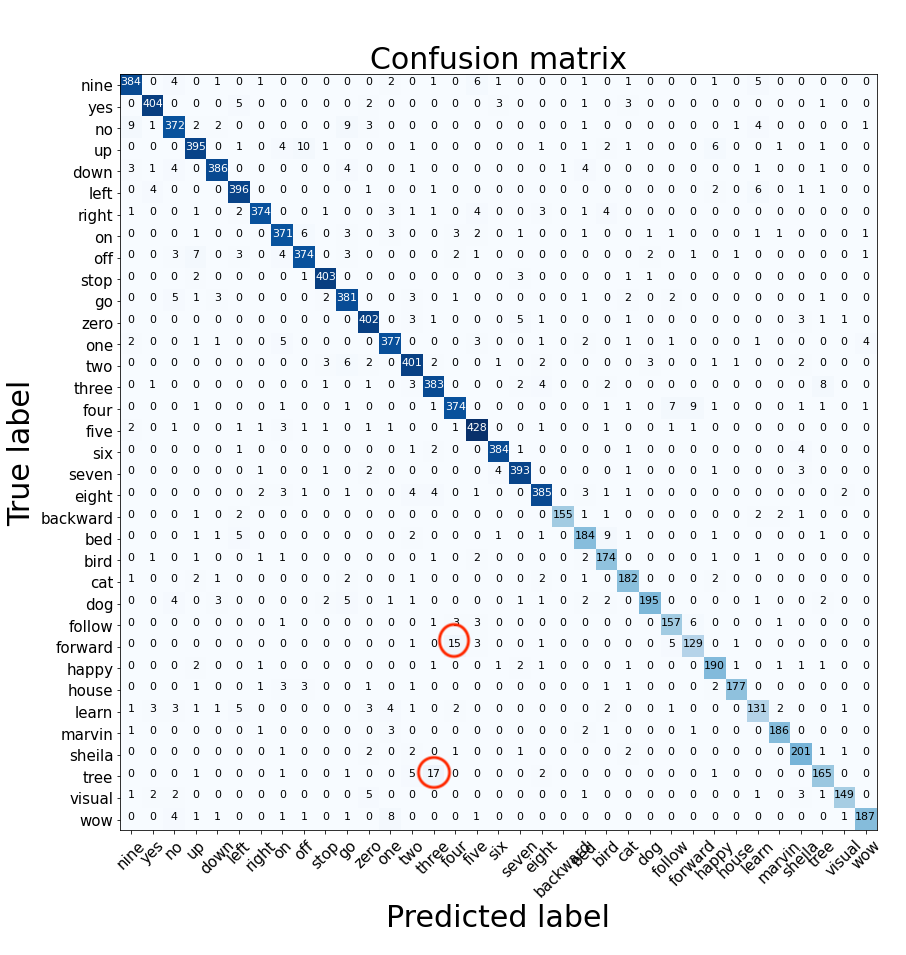}
    \caption{Confusion Matrix (Attention RNN) }
    \label{fig:cm}
\end{figure}

Figure \ref{fig:cm} shows the results of the attention RNN model in the form of a confusion matrix. The highlighted blue diagonal shows the correctly predicted keywords. The numbers around the diagonal show the true labels against the incorrectly predicted labels. We are able to see that the number of incorrect predictions are high (some examples circled in red) when the commands have similar utterances, such as 'tree' and 'three' or when the words begin with the same speech sound (phonemes) such as 'forward' and 'four'. For such errors, a possible solution would be to add 'context' into our model to enable it to understand the difference between misleading word pairs or groups. 

\section{Conclusion}

Speech Recognition has been a field of extensive work due to its importance in human-machine interaction. From the earliest implementation of HMMs that viewed a speech signal as a piece-wise stationary signal to the more recent more attractive acoustic modeling approach of using framework such as CNNs and RNNs, speech recognition has been an extensively researched field. In this study, we are able to successfully run multiple such models with varying architectures on a publicly available and widely used dataset and conduct a comparative analysis to understand which frameworks achieve the highest performance for commands detection.

In our experiments, we find that deep neural nets can significantly improve the performance compared with the Hidden Markov Model with Gaussian Mixture. Among the deep architectures, we notice that RNNs, the model that considers the sequence order of the inputs, are more accurate than the CNN which does not take the position of the input into account. Of the two RNN models that we explore, we find that RNN with BiLSTM and Attention mechanism achieves the highest accuracy of 93.9 \%. However, it is only a trivial improvement over RNN with unidirectional LSTM, which is a much simpler model. Therefore, we assume that although attention has been shown to improve the performance significantly in sequence-to-sequence (encoder-decoder) models, it does not contribute much in our task.

\subsection{Future Work}
For future works, it is worth investigating additional models on this task such as acoustic HMMs and some combination of existing frameworks such as DenseNet + BiLSTM for boosting accuracy. It is also interesting to explore more complex data augmentation mechanisms in audio samples. Similarly, in the field of Natural Language Processing, using pre-trained models has become the 'advocated' approach with BERT producing state-of-the-art results for a variety of downstream NLP tasks. It is worth trying the BERT and fine-tune the weights on this task.

Our study can also be extended to incorporate speech recognition for sentences. By stacking pairs of words, we can detect more complex commands and use sequence-to-sequence techniques for modeling natural language conversations.

\clearpage

\renewcommand\bibname{References}
\bibliographystyle{abbrv}
\bibliography{bib}

\section{Contributions}

All team members contribute to the data cleaning, EDA, modeling, and report writing processes. TL built the HMMGMM and RNN with unidirectional LSTM. BL implemented the CNN model. SR built the RNN with BiLSTM and Attention.  

\section{GitHub Link}
https://github.com/sumedharai12/DetectingSpeechCommands

\end{document}